\documentclass[a4paper,twoside]{article}

\usepackage{epsfig}
\usepackage{subcaption}
\usepackage{calc}
\usepackage{amssymb}
\usepackage{amstext}
\usepackage{amsmath}
\usepackage{amsthm}
\usepackage{multicol}
\usepackage{pslatex}
\usepackage{apalike}
\usepackage{booktabs}
\usepackage[bottom]{footmisc}
\usepackage{SCITEPRESS}   

\begin{document}
\title{Student teacher interaction while learning computer science. Early results from an experiment on undergraduates.}
%

\author{\authorname{Manuela-Andreea Petrescu\sup{1}\orcidAuthor{0000-0002-9537-1466}, Kuderna-Iulian Benta{\sup{2}}\orcidAuthor{0000-0002-4245-599}}
\affiliation{\sup{1} Department of Computer Science, Faculty of Mathematics and Computer Science, Babe\c{s}-Bolyai University, Cluj-Napoca, Romania}
\email{manuela.petrescu@ubbcluj.ro, kuderna.benta@ubbcluj.ro}
}


%

\abstract{
The scope of this paper was to find out how the students in Computer Science perceive different teaching styles and how the teaching style impacts the learning desire and interest in the course. To find out, we designed and implemented an experiment in which the same groups of students (86 students) were exposed to different teaching styles (presented by the same teacher at a difference of two weeks between lectures). We tried to minimize external factors' impact by carefully selecting the dates (close ones), having the courses in the same classroom and on the same day of the week, at the same hour, and checking the number and the complexity of the introduced items to be comparable. We asked for students' feedback and we define a set of countable body signs for their involvement in the course. The results were comparable by both metrics (body language) and text analysis results, students prefer a more interactive course, with a relaxing atmosphere, and are keener to learn in these conditions.}
\onecolumn \maketitle \normalsize \setcounter{footnote}{0} \vfill

\section{Introduction}

We propose to validate the importance of interaction, general atmosphere, and teacher’s passion in the learning process in Computer Science courses. We created an experiment where the delivered information is comparable but the teaching style and interaction are deliberately modified. 

By analyzing the student’s perception - both on visual level/observed behavior  (they were sleepy, yawns, laying on the benches) and from their collected textual feedback. The feedback was analyzed manually using thematic analysis.

We wanted to find out if there is a difference in perceived understanding and interest from 
students when we use interactive methods combined with elements of nonverbal communication compared to a classical teaching method (just presenting the information). Also, to find out if the teaching methodology impacted students' perception related to lecture in terms of affectiveness (they feel enthusiastic, good, interested or bored, discouraged) and in cognitive terms (they understand better / worse). We created a list of nonverbal elements (yawns, laying on the desks, arms positions, and so on) that we analyzed during the lectures and analyzed their frequency. At the same time, we asked the students for anonymous feedback. 

Education is a key component of society, influencing a country's future development; due to its importance, countries allocate a part of their expense budget to the educational system. Except for money, other factors influence the attractiveness of a specific domain. For Computer Science and Mathematics domain, for example, the interests of secondary schools students depend on a set of factors: the student's socioeconomic status, performance, self-efficacy, motivation, engagement, and task value beliefs \cite{Kahraman22,Spieler20}. Universities noticed that the number of students that graduate is smaller compared to the enrolled students, some tried to decrease the drop-out rates by offering additional materials and offering some courses in an online or hybrid format to minimize the tuition fees and to increase accessibility. Online courses increase accessibility as the students can learn whenever they have time, they can learn at their own pace, and having access to resources before the course allows them to ask more and more complex questions \cite{Baquerizo20}; however neither going online solved all the problems. Paper \cite{Baquerizo20} analyzes methods to motivate students in an online environment, as \cite{Petrescu22} mentioned students' difficulty in finding a quiet place for learning.

\section {Literature Review}
\label{sec:literature_review}
  
A key to success for companies and countries is human capital, well prepared, capable people can raise a company or a country. To have prepared and well-qualified people, education, the methods to transmit information, and information accessibility play a major role. Due to this need, different methods were analyzed and proposed in the literature with the declared scope to increase teaching effectiveness, some of the methods were strictly related to the computer science domain \cite{Liu2022,Salas2017,Marcus2020}. Teaching methods in computer science were also subjects for other papers/books that analyzed the impact of teaching styles, some mention teacher collaboration methods, technical professional development \cite{Wang21}, or offering guides for computer science instructors in universities \cite{Hazzan2020}.

In \cite{DBLP:conf/csedu/ErascuM22} a study took place while applying Student-centered Learning in Computer Science. One of the three research questions was trying to investigate if the student centered approach is perceived as better compared to a control group. Pre-tests and post-tests were used. The results show a positive impact of the student centered approach on the active learning dimension. However, our study is oriented more toward the subjective perception of the didactic process. The perspective from \cite{Makhlouf21} is similar to ours in the sense that the comments are clustered in a small number of classes with similar meaning to build up an automated feedback system that facilitates individualized feedback given by the professor to students. By contrast our work focus on understanding the impact of nonverbal cues effect on the learning process.

The relationship between emotions and engagement in learning in an e-learning controlled environment is piloted using three biosensors \cite{Khaled18}: a camera, a 14-channel EEG, and an eye-tracker. The lab-like setup, the reduced number of participants (15), and the small number of emotions considered give us inspiration for what we could use while measuring the learning engagement level in real-time to improve the quality of teaching. Complementary to our approach in \cite{Zeyad21} an automatic authoring tool based on the Socratic dialog is proposed to improve code comprehension. The authors claim that it improves students’ programming knowledge by 43\% in terms of learning gain.

\section{Design of the Study/Experiment}
\label{sec:design}

\subsection{ Experimental Setup}
The experiment involved a Computer Science teacher from a university and students from Computer Science. The courses were lectured in English during the laboratory, it was a first-year course called Computer System Architecture and was structured in two parts: a theoretical part where different teaching methods were implemented that lasted for an hour, and the second hour was preserved for exercises and examples and a QA (Question and Answer) session. Because we considered the students to understand the theoretical part better, all the introduced notions had small examples during the theoretical part and when possible, there were mentioned correlations with previous knowledge/information the students should have acquired. The courses were held in the same location at the same hour to minimize the environmental impact on the student's perception. Both courses were presented using a projector for a presentation, and in the first one, the blackboard was also used to present other examples.
The weather or some specific events could impact the student’s perception, but we choose to have the course as close to one another as possible (two weeks difference). We were fortunate that there were no major weather differences during the period. We selected a time of year that did not have specific events (holidays, the first days of the semester, concerts in the city, and so on).
We organized the first course to be the interactive one and the second to be in a classic "lecturing" style, because we asked for feedback after each course, we wanted them to be able to analyze the differences ("lecturing" style in our opinion is more commonly used than the interactive style, where the teacher must do additional efforts to imply the students).

\subsection{Participants}
 86 students from Computer Science participated in the experiment, 75 of them were students aleatory selected from the first year, and 11 students were in the second year of study that did not pass the course exam in the first year and had to retake the course. They opted to come to a specific laboratory based on a timetable and to have the lab with a preferred teacher. There was no specific selection depending on gender or other criteria, the participant set was randomly selected. The participants were informed about the experiment, and that their participation was optional (that’s why the number of students that provided feedback and participated actively in the survey by providing answers is lower than the overall number of students that took part in the lectures). The students were grouped into five groups, each group having between 15 and 17 people. For each group was delivered Lecture 1 and Lecture 2.
 
 The participants were required to provide:
 1. Feedback after each lecture using an anonymous form that contained questions related to how well they understood the concepts and the presented information, 
 2. Give their opinion about the delivery in terms of interaction, voice volume, body language, the overall atmosphere, and if the delivery had an impact on the interest in learning in that specific domain. We decided to have the same participants set for both lectures/setup so they could appreciate and compare the teaching styles and the teaching style's impact on them because an event or a piece of information could have a different impact depending on the involved person's characteristics and mood. 
 
 \subsection{Lecture comparison in terms of new information}
To have a valid experiment, and to be able to correctly compare the results of the teaching methods, we selected two introductory courses: \textit{Conversions and Complementary Code} and \textit{ASM Arithmetic expressions} that were similar in terms of presented information (content, difficulty, and the number of new notions). We analyzed the presented information and structured it into two categories: \textbf{Build-up} and \textbf{New Info}. In each category, we added the topics that were discussed and we assigned a difficulty level (DF) from 1 to 3, where 1 is easy, 2 is medium and 3 represents a difficult topic. We summarized the points for each topic to find out what was the overall course difficulty. When we assigned the difficulty level we also took into consideration if a topic was a completely \textbf{new} one or if it had an increased complexity \textbf{(build-up)} information for a subject that the students already should know. 
We summarized the difficulty levels (DF) and the results were close (12 to 13) for new information and (1 to 3) for build-up information. We conclude that the information presented had a comparable degree of difficulty and a comparable number of items. Below we visualized in Table \ref{table:comp} the topics presented in each lecture, the build-up type is noted with \textit{''b''}. 
 \begin{table}[!htp]
  \caption{New \& build-up topics comparison}
  \label{table:comp} 
  \scriptsize
  \begin{tabular}{|p{2.1cm}|p{0.2cm}|p{0.3cm}|p{2cm}|p{0.2cm}|p{0.3cm}|}
    \hline
    \textbf{New info - L1} & \textbf{DF} & \textbf{Type} & \textbf{New info - L2} & \textbf{DF} & \textbf{Type} \\
    \hline
     Conversion to base 16 and viceversa & 1 &  b & ADC \& SBB instructions & 2 & b\\ \hline
     Conversion to base 8 and viceversa & 1 & b & IDIV \& IMUL & 2 & b \\   \hline
     Addition \& Substraction base 2 & 1 & b & CBW \& CWD & 2 & b \\ \hline
     Addition \& Substraction base 16 & 1 & b & CWDE \& CDQ & 2 & b  \\ \hline
     Sign bit & 1 & new & Declaring variables & 1 & new    \\ \hline
     Complementary code & 2 & new & Declaring constants & 1 & new  \\ \hline
     Complement to 2 & 2 & new & Push / Pop flags \& registries & 2 & new    \\ \hline
     Representation size & 1 & new & Push / Pop flags \& registries & 2 & new     \\ \hline
     Asm tools \&  Program example & 2 & new & Stack applications & 1 & new \\ \hline
\end{tabular}
\end{table}

\section{Data Collection and Analysis}
\label{sec:data_collection}

The responses were collected anonymously in the form of open answers, and the students were informed related to the purpose of the questions and also about how their responses will be used. The students were asked to provide optional feedback on a quiz at the end, but there was no time limit. The response quiz remained open for two weeks. After this interval, we considered that we will not get other valid responses. However, most of the responses (93\%) were sent on the same day - we collected the timestamp by checking the response poll at different intervals. 

We opted for open questions as they offer a better and more profound understanding. We used quantitative methods and more specific questionnaire surveys as they were defined in the empirical community standards \cite{ACM}. These methods were previously used in Computer Science related studies \cite{Tichy95,Redmond13,Borza22}. For text interpreting, we used thematic analysis \cite{Braun19} to interpret the text and take into account the recommendations mentioned in \cite{Kiger20} for free text interpretation. The method was previously used in Software Engineering in other studies: \cite{Cruzes11,Peggy15}. The reflexive approach was applied as described by \cite{Kiger20,enase21} 



A number of 86 students participated in the experiment, and their feedback was monitored and checked in two ways:
\begin{itemize}
    \item By participating voluntarily  and providing answers for a quiz  - we gathered 73 answers.
    \item By observing their body language at the end and during the lectures and looking for quantifiable signs of their interest such as the number of yawns, the number of questions asked, and for less quantifiable signs such as laying on the desks, body position, eye movement, and so on.
\end{itemize}

We asked the following open questions related to each course to get more information: \textit{\textbf{Q1:} Did you understand the information presented in the lecture?},
\textit{\textbf{Q2:}  How was the delivery of the course and the teaching style?}
and  \textit{\textbf{Q3:} What was the effect of the teaching style related to learning interest? What did you like/ dislike about this course?} Next, we will break down and analyze each research question separately.

\subsection{Q1:Did you understand the information presented in the lecture? }
We tried to structure the information in a logical and easy-to-understand way, where one presented topic is tightly related to the previous one, thus creating an easy-to-follow and clear presentation. The presented information was doubled by exercises (9 answers in total appreciated integrating exercises in the theoretical part): \textit{"we got involved in the exercises too"}. In the answers provided for the course structure and content, we classified the keywords into two classes, one related to content and structure and one related to the content's applicability Table \ref{tab:2 }. 

\begin{table}[!htp]\centering
\caption{Keywords class}\label{tab:2 }
\scriptsize
\begin{tabular}{|p{1.2cm}|p{5.5cm}|}
\hline
\toprule
\textbf{Item Class} &\textbf{Selected keywords}
\\\midrule
\hline
Structure & Well structured, Easy to understand, Clear, On the point, Good, Exercises \\
\hline
Applicability & Useful, Valuable, Changed the perspective, \\
\bottomrule
\hline
\end{tabular}
\end{table}

We analyzed the prevalence of the keywords in the received answers, for each course, each student provided three answers related to structure and content, delivery, and effect. Sometimes content related keywords appeared in two out of three, or even in all three answers, sometimes more keywords appeared in one answer. We measured them to establish the general impact of a feature and then we compared the results for each lecture. We analyzed the answers, each text could contain none, one or more keywords (clearly, well structured, and so on) that were related to a positive appreciation. When we summarized all the keywords, they were more than the number of the students, so the percentage compared to the number of the students was 114.58\%. Same method was performed when analyzing the responses for the second lecture and we obtained 74.07\%. The positive appreciation related to content and structure was higher in the first course compared to the second course (114.58\% vs 74.07\%). We took into consideration two factors: in the first course the students were more motivated to provide longer answers, thus increasing the prevalence of specific keywords (number of words/answer/course), and having 74.07\% of answers appreciating positively the structure and the content in the second course, we considered that both courses scored high in terms of content and structure, i.e.  \textit{"everything was explained clearly and if I had any questions they were answered"},vs. \textit{"The information was structured well, but the delivery is not as helpful compared to more active and interactive one".}

However, the keyword reflecting the overall applicability was much higher in the first course, we believe the difference lies in the information taught: in the first lecture the students found out why we have specific data types, information generally useful \textit{"The content was interesting and it changed the way I view computers and programming"}. In the second lecture, they found only instructions related to Assembly Programming Language, and considered that the topics presented in lecture 2 do not have any applicability.   

 \begin{figure}[htbp!]
    \includegraphics[width=0.48\textwidth]{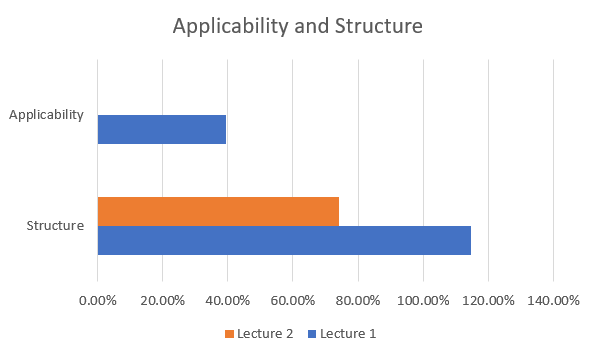}
    \caption{Applicability and Structure Keywords Prevalence}
    \label{fig:q2Class}
\end{figure}

Based on their answers, we concluded that:\newline
\textbf{A.} The information was structured clearly and easily. \newline
\textbf{B.} The students appreciated the applicability of the presented information in the first course, thus reinforcing the fact that students are more interested in practical aspects (when they realize the theoretical part's applicability).

\subsection{Q2: How was the delivery of the course and the teaching style?}

For the second question, we identified two major classes: one that refers to the delivery itself containing keywords such as pace, interactivity, and the volume of the voice, and one that refers to the teacher's involvement: passionate, calm, open-minded. We counted the prevalence of the first class in the answers for the first lecture (35 mentions), the relatively similar prevalence was obtained for the second lecture, for the same class, but the appreciation had a negative connotation (31 mentions): no interaction, blunt, boring, bad delivery. In both cases there were key items with  neutral connotation: delivered straight, or \textit{"Okay"}. The Table \ref{table:q2Class} reflects the resulting keywords for the second question.

\begin{table}[!htp]\centering
\caption{Delivery Keywords Prevalence}\label{table:q2Class}
\scriptsize
\begin{tabular}{|p{1.8cm}|p{5.0cm}|}
\hline\toprule
\textbf{Item Class} &\textbf{Selected keywords (number of appearance)} \\\hline\midrule
Content Delivery Lecture 1 &Interactive / dynamic (9), Well delivered (9), Right pace (6), Presentation ok /good (4), Open discussion (2), Check understanding (2), Answers (2), PPT and white board (1). \\\hline
Content Delivery Lecture 2 &No interaction/interactivity (19), Pace too fast (5), Blunt/Boring (4), Ok (4), Bad delivery (1), Monotone (1), Delivered straight (1) \\\hline
Teacher’s attitude Lecture 1 &Relaxed Atmosphere (9), Friendly (5), Active (4), Calm (3), Cherry attitude (1), Nice (1), Helpful (2), Funny (2), , Involved (2), Open-minded (1) \\\hline
Teacher’s attitude Lecture 2 &Cold and distant way (2), No emphasis (1) \\
\bottomrule\hline
\end{tabular}
\end{table}

The answers from the first lecture provided information not only about the delivery but also their perception related to a supportive environment: \textit{"the information was delivered relatively good, you made sure we were not behind"}. We also got short answers related more to the lecture difficulty: \textit{"Pretty basic stuff, she made it feel interesting though"}. 
In the second lecture, there were still examples presented, but the delivery changed: sitting down, fewer body movements and gestures, less eye contact, and so on. Some students managed to figure out some of the teacher's delivery methods that influenced the presentation: \textit{"The information was hard to follow. There was no emphasis put on the more important information"}, others stated that \textit{"I managed to understand the information in its entirety. The content in itself was logical and the required explanation or clarification of the written theory was provided by the professor".}

The second class keywords that were identified can be traced more to the lecturer's personality than to the teaching style: active, friendly, calm, helpful, and even funny: \textit{"The atmosphere was relaxed and the course was taught in a fun way"}

It was interesting that all these key items appeared in the answers from the first course, and none appeared in the answers related to the second course. There were no key items reflecting personality characteristics. However, key items were referring to the overall presentation \textit{"It was bland, boring, but I managed to understand"}. 

For this section, we can conclude that: \newline
\textbf{A.} The presentation mode influenced the perceived difficulty level of the presented information. The students found it more difficult to concentrate and not lose focus during the course when the course was not presented in an interactive way (even if there were exercises and examples and the teacher answered the questions). \newline
\textbf{B.} Students reacted positively to the lecture's personality, appreciating a more "open discussion" lecture than a classical one. In the classical \textit{"just words"} style (as one of the students mentioned", the students were bored \textit{" it felt like a YouTube tutorial, tutorials are useful but not enjoyable"}.

\subsection{Q3: What was the effect of the teaching style related to learning interest? What did you like/ dislike about this course?}

We expected that teaching style to have an impact on the overall interest. We analyzed the teaching style effect on students by two methods: 
\begin{itemize}
    \item We analyzed their body language during and at the end of the course to look for signs. 
    \item We asked for their feedback in the form of open questions.
\end{itemize}{}

In the interactive lecture, when they were actively asked to answer questions and constantly provoked related to their knowledge or to make logical deductions, the students were active, we counted a small number of yawns only at the beginning of the courses that took place from 8 am. In the other course, fatigue, sleepiness, and uninterested signs of body language appeared  approximately half an hour after the course started, and much faster (after 10-15 minutes) in the courses that took place at 8 am. Next, in Table \ref{tab:4} can be seen a comparison between the observed body language signs, two lectures took place in the 8 am - 10 am interval, two lectures took place in the 10 am-12 am interval, and one lecture took place at 12 am - 2 pm interval for each topic. So in total, we analyzed 10h lectures, by observing and counting how many times a specific behavior appeared.

\begin{table}[!htp]\centering
\caption{Body language signs observed during the lectures}\label{tab:4}
\tiny
\begin{tabular}{|p{1.1cm}|p{1.1cm}|p{1.1cm}|p{1.1cm}|p{1.1cm}|}\hline\toprule
\textbf{Body Signs That Appeared (per lecture)} &\textbf{L1 (8.00-10.00)} &\textbf{L1 (10.00-12.00)} &\textbf{L2 (8.00-10.00)} &\textbf{L2 (10.00-12.00)} \\\midrule\hline
Laying on the back, related positions &Only in the first 10-15 min, 22\% from students &No signs &After first 10-15 min, at 68\% from students &After first 30 min, at 64\% from students \\\hline
Yawns &Only in the first 15 min (9\% of students) &No yawns &Appeared at 30\% of students, more than once &Appeared at 25\% of students, more than once \\\hline
Unfocused eye look &Only in the first 10-15min, at 13\% of students &No signs noticed &Appeared to 90\% from students &After 30 min, at 85\% of students \\\hline
\bottomrule\hline
\end{tabular}
\end{table}

We also performed a text analysis by manually processing the students' responses to the open questions, and the results were congruent with the body language that was noticed during the lectures. Most of the students appreciated the interactive lecture, and some of them mentioned elements of interactivity: \textit{"I enjoyed the delivery and the fact that the class was interactive. The presentation gave everything a bit more structure"}. Some answers reflected more the effect, even mentioning causes that influenced the learning environment: \textit{"I feel really good at this lab because the atmosphere is really friendly and open-minded. I really enjoy it."}. 

However, no charm and no teaching method can overcome the call of nature, in the course that took place during the lunch period, 20\% of them mentioned they were hungry: \textit{"I want to learn more and I'm also hungry.."}, we even had a response that mentioned: \textit{"Really dizzy, hungover from last night party"}. Curiously, there were no such mentions for the second course (which was held during the same hours). We assumed based on the shortness of answers that they just decided to provide less information. 

As for the second part of the experiment, the students mentioned that they were  bored and missed the interest, we had short, neutral answers: \textit{"Presented in a linear way", "It was okay"}, or little longer answers: \textit{"Monotone, kind of boring, not interactive"}. Because they participated in both courses (same teacher, different teaching styles), they performed a comparison of the teaching styles, thus offering reliable information related to the effect of the teaching style. They realized the difference and emit conclusions based on their experience: \textit{"I lost my concentration level in the middle of the course and I missed the interaction with the teacher"} and even recognized doing completely something else: \textit{"I disliked the teaching style, at some points I even realized I completely zoned out and just worked on other stuff on my laptop, completely ignoring the teacher."} Item classes are detailed in the next table.

\begin{table}[htp!]\centering
\scriptsize
\label{bodyLang1}
\caption{Classes of items}\label{tab:5 }
\begin{tabular}{|p{1.2cm}|p{5.0cm}|}
\hline
Needs extra work & \textit{"need more time, have to reread, need to learn more, information was difficult, less clear"}\\
\hline
Easy & \textit{"easy to understand, easy, understand all, not overwhelming"}\\
\hline
Environment & \textit{"less stressed, relaxed atmosphere, monotonous, boring, attention not drawn
"} \\
\hline
Student's perception & \textit{"entertained, liked, enjoy, surprised, no fear, no boring"} or \textit{"did not like, felt discouraging, not pay attention"} \\
\hline
Student's reaction &  \textit{"waiting for next courses, interested", "impact learning desire", "worried about course difficulty"} \\
\hline

\hline
\end{tabular}
\end{table}

\begin{figure}[htbp!]
    \includegraphics[width=0.48\textwidth]{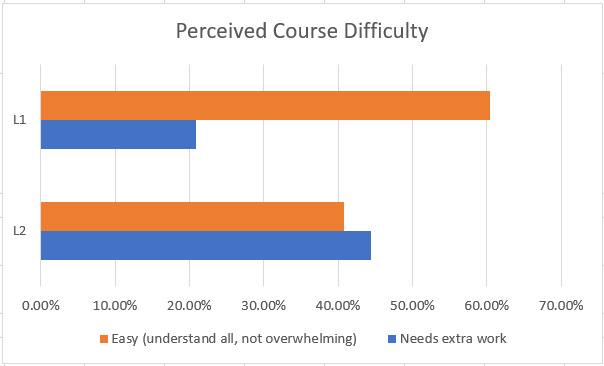}
    \caption{Perceived Course Difficulty }
    \label{fig:CourseDif}
\end{figure}

To see how much the teaching style impacted how the students perceived the difficulty (even if the topics had a similar level of difficulty), we checked how many students reported the information to be easy for the first course compared to the second, 60.42\% of the answers considered the information delivered in the first course to be "easy" compared to 40.74\% of the answers for the second course. The numbers correlate with the number of answers that mentioned: "extra work" is needed: 20.83\% for the first course compared to 44.44\% for the second course: Fig \ref{fig:CourseDif}.

\begin{figure}[htbp!]
    \includegraphics[width=0.5\textwidth]{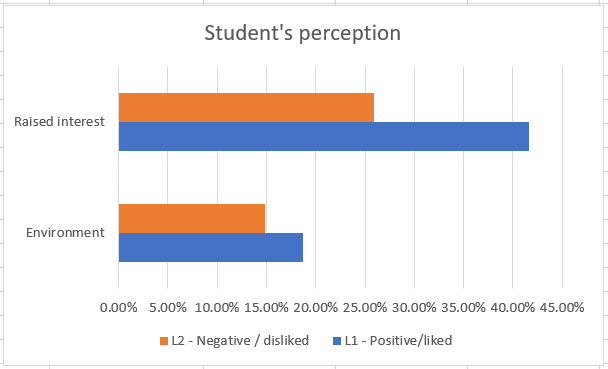}
    \caption{Student's Perception}
    \label{fig:StudPerc}
\end{figure}

Regarding the environment and the course atmosphere, positive characterizations were received for the first lecture \textit{"relaxed atmosphere, less stressed"}, which appeared in 18.75\% of the answers for the first course. Negative characterizations were received for the second course \textit{"monotonous, boring"} (14.81\% of the total number of answers).41.67\% of the answers for the first course mentioned the students \textit{"liked, enjoyed, entertained"} as 25.93\% of the answers from the second course mentioned \textit{"did not like, felt discouraging"}, Fig \ref{fig:StudPerc}. Moreover, 18.75\% of the answers in the first course mentioned that they are \textit{"intrigued to learn more"} even if they are worried about the course complexity (6.25\% of the answers). In the second course, 37.04\% of answers mentioned that the teaching style impacted their learning desire: \textit{"in time I think it would impact my learning desire.","The switch in teaching style felt discouraging compared to previous labs", "It impacted the learning desire quite hard, making me not paying attention, and lose focus"}.

Based on their answers, we concluded that: \newline
\textbf{A.} The teaching style impacted their attention and as a consequence, their desire to learn.  \newline
\textbf{B.} The teacher's ability to relate and to create a relaxing calm environment is mentioned by 18.75\% of them, so this factor is important for a large majority. \newline
\textbf{C.} There might be a problem with the overall student-teacher relationship in the high-school cycle as several students appreciated the fact that \textit{"the questions were answered", "we could ask without FEAR", and "open talk discussion"},  all these suggest a more profound problem. \newline
\textbf{D.} No teaching style and no abilities can come against nature's call (hungry, sleepiness, and so on).


\section{Threats to Validity}
\label{sec:ThreatsToValidity}

According to ACM (Association for Computing Machinery) standards for software engineering research \cite{ACM}, we were aware that we need to address possible threats to validity that could impact the obtained results and we decided to analyze the following: research ethics, target population and participant selection, drop-out measures, environmental threats, methods to decrease the subjective form of data processing, hypotheses are missing. 

Research ethics: At the beginning of the experiment (at each course) the students were informed that they will take part in an experiment that consists in teaching two courses in two different styles, and that their participation in the survey is optional and anonymous. We also informed them about the scope of the study and how the collected data will be used.

Target population and participant selection: The groups are formed in alphabetical order based on the student's surname, there were 5 aleatory groups selected for the experiment (from a total of 14 groups of students enrolled in Computer Science - English line), and all of them were required to participate in the study. Thus, we assured that the selection was aleatory.

Drop-out measures: As being optional, a number of students decided not to participate, and they did not provide answers; we did not want to use any constraint methods, and we could not influence the drop-out rates.

Environmental threats or inappropriate design for the conditions under which the experiment took place: human subjects can be influenced in many ways, that's why we wanted to control as many external factors as possible: we used the same classrooms, the same times of day, and close dates. We used the same materials (video projector), and we took into consideration course difficulty (to same a similar grade of difficulty in terms of new information and complexity). All these factors were taken into consideration so their perception/opinion was influenced only by different teaching styles.

Data processing: the process used for data processing was in concordance with defined processes and was also used in other computer science papers. We defined two distinct methods to validate the results, by observing behavior in a countable manner (counting the number of yawns) and by having a text analysis of the received responses.

Hypotheses are missing: We assumed that teaching style and teacher-students relation impacts the learning desire even in a Computer Science course. 

\section{Conclusions and future work}
\label{sec:conclusion}

We wanted to find out how the students perceive different teaching styles and how the teaching style impacts the learning desire and interest in the course. To be able to have valid data, we used the same environment (classes, course hours, close dates), and the same students participated and were exposed to both teaching styles, thus being able to evaluate and compare them. We paid attention to the introduced concepts, they were comparable in terms of number and complexity. We tried to take into consideration all the aspects that could influence the outcome and could become a threat to the validity of the results. We analyzed the effects and the results of the teaching styles in two methods - the first one: asking for students' feedback after each course and the second one: defining a set of countable behavioral signs (yawns, laying on the bench). Both methods returned the same result: students were more attracted to the interactive teaching style and did not show behavioral signs of boredom. Teacher behavior manager to create a "calm and relaxing atmosphere" in the first course. In the second course, when the teaching style was deliberately changed, to a non-interactive one, the signs of boredom could be measured and the students reported a lack of interest, daydreaming, and even doing a completely different thing. 
A more rigorous comparison by using different natural language processing techniques could be used to show the correlation between the two performed analyses. We plan to define a metric to measure more accurately the student's interest related to the teaching style and also to replicate the experiment for a whole semester so we can evaluate the results using exam outcomes. A multi-modal affective state monitoring could be designed on the class level to pervasively measure students' emotions and mood while learning using wearable devices \cite{Benta15}.








{\bf Funding}
The publication of this article was supported by the 2022 Development Fund of the Babe\c{s}-Bolyai University.

\bibliographystyle{apalike}
{\small
\bibliography{asc}}

\end{document}